\documentclass[superscriptaddress, amsmath,amssymb,preprint,showkeys,showpacs]{revtex4-1}

\usepackage{graphicx}
\usepackage{graphics}
\usepackage{amssymb}
\usepackage{amsmath,bm}
\usepackage{float}
\usepackage[usenames,dvipsnames]{color}
\usepackage[export]{adjustbox}
\usepackage{hyperref}
\begin{document}

\title{Near-field measurement of modal interference in optical nanofibers for sub-Angstrom radius sensitivity}

\author{Fredrik K. Fatemi}
\affiliation{Army Research Laboratory, Adelphi, MD 20783, USA}
\email{Corresponding author: fredrik.k.fatemi.civ@mail.mil}
\author{Jonathan E. Hoffman}
\affiliation{Joint Quantum Institute, Department of Physics, University of Maryland and National Institute of Standards and Technology, College Park, MD 20742, USA.}
\author{Pablo Solano}
\affiliation{Joint Quantum Institute, Department of Physics, University of Maryland and National Institute of Standards and Technology, College Park, MD 20742, USA.}
\author{Eliot F. Fenton}
\affiliation{Joint Quantum Institute, Department of Physics, University of Maryland and National Institute of Standards and Technology, College Park, MD 20742, USA.}
\author{Guy Beadie}
\affiliation{Naval Research Laboratory, Washington DC, 20375, USA.}
\author{Steven L. Rolston}
\affiliation{Joint Quantum Institute, Department of Physics, University of Maryland and National Institute of Standards and Technology, College Park, MD 20742, USA.}
\author{Luis A. Orozco}
\affiliation{Joint Quantum Institute, Department of Physics, University of Maryland and National Institute of Standards and Technology, College Park, MD 20742, USA.}

\date{\today}

\begin{abstract}
Optical nanofibers (ONF) of subwavelength dimensions confine light in modes with a strong evanescent field that can trap, probe, and manipulate nearby quantum systems.  To measure the evanescent field and propagating modes, and to optimize ONF performance, a surface probe is desirable during fabrication. We demonstrate a nondestructive measurement of light propagation in ONFs by sampling the local evanescent field with a microfiber.  This approach reveals the behavior of all propagating modes, and because the modal beat lengths in cylindrical waveguides depend strongly on radius, simultaneously provides exquisite sensitivity to the ONF radius.  We show that our measured spatial frequencies provide a map of the average ONF radius (over a 600 micrometer window) along the 10 mm ONF waist with 40 picometer resolution and high signal-to-noise ratio.  The measurements agree with scanning electron microscopy (SEM) to within SEM instrument resolution.  This fast method is immune to polarization, intrinsic birefringence, mechanical vibrations, scattered light, and provides a set of constraints to protect from systematic errors in the measurements
\end{abstract}

\maketitle

\section{Introduction}

The evanescent fields outside ONFs~\cite{Tong2003} allow strong interactions with the surrounding medium~\cite{LeKien2005,Morrissey2013,Yalla2012,Fujiwara2011,Arnodispersive,Vetsch2010}.  These waveguides have enabled several advances in quantum information technologies, including optical switches~\cite{OShea2013} and atom-mediated optical isolators~\cite{Mitsch_NC, Petersen2014}, but have also facilitated fundamental experiments in nonlinear atom-light interactions~\cite{Pati}, atom-number-resolving detection~\cite{Polzik}, electromagnetically induced transparency \cite{Sayrin2015,Gouraud2015,Jones2015,Kumar15}, and Bragg reflection from atoms \cite{sorensen16,Corzo16}. 

The control achieved in fabrication of high transmission ONFs in both the fundamental mode~\cite{Hoffman2014} and higher-order modes (HOM)~\cite{Ravets2013a} is excellent.  A nondestructive tool to accurately measure the ONF radius and characterize the propagation of the modes is crucial for optimizing both fundamental and applied uses of ONFs. For example, the ONF radius, $R$, governs the coupling between a nearby atom (trapped or free) and the allowed nanofiber modes (see \textit{e.g.} Ref.~\cite{LeKien2005}).  These modes have distinct effective refractive indices ($n_{\rm{eff}}$) that depend strongly on $R$. 
We can use modal beating to extract both propagation properties and ONF geometry for optimizing atom-photon coupling.  We recently examined propagation characteristics by imaging Rayleigh scattered (RS) light~\cite{Fatemi_optica}.  While RS is an effective tool for analyzing some propagation behavior, it uses far-field imaging and is hampered by excess scattered light, mechanical vibrations of the ONF, and imaging resolution so that only long-period modal beating can be cleanly resolved~\cite{Fatemi_optica,Szczurowski}.  

To circumvent the limitations of far-field imaging, contact techniques between an ONF and a microfiber probe have previously been used to measure the local radius variations of an ONF with sub-nanometer precision by propagating light through the microfiber probe and observing the mode spectrum of a whispering gallery or a composite photonic crystal cavity~\cite{Birks, Sumetsky,Semenova15,Keloth15}. Though these approaches accurately measure ONF dimensions, they do not measure the propagation characteristics of the ONF.  Other contact techniques rely on changes in the amplitude of the transmitted light, becoming sensitive to polarization and van der Waals forces, as in the recent work by Madsen {\it et al.}~\cite{Madsen2016}, but do not directly sample the modal composition of the local field.

In this work, we measure the beat lengths between propagating nanofiber modes in the near field over the ONF length through evanescent coupling to a microfiber probe.  This near-field approach provides high ($\approx$1$~\mu$m) longitudinal resolution capable of measuring beating between any mode pairs and is immune to excess scattered light.  The spatial frequencies are used to map the mean of the ONF radius to within 40 picometers over 600 $\mu$m measurement windows along the ONF taper and waist. The amplitude of the beat frequency only enters in the signal to noise ratio of the resonance, which in this technique can be orders of magnitude better than those we obtained with Rayleigh scattering \cite{Fatemi_optica}. 

\section{Theoretical Considerations}\label{subsec:RSMtheory}
A standard optical fiber consists of a core of refractive index $n_\mathrm{core}$ and radius $a$, surrounded by a cladding with lower refractive index $n_\mathrm{clad}$ and radius $R$.  In our ONFs, $R$ is reduced to subwavelength dimensions by a flame-brush technique~\cite{Hoffman2014}.  At these dimensions, the ONF can be considered as a simple dielectric of index $n_\mathrm{ONF}$ = $n_\mathrm{clad}$, surrounded by $n$ = 1.0, as the original fiber core becomes negligible.  The tapers connecting the standard fiber on the input and output side to the ONF waist have milliradian angles for adiabatic propagation~\cite{Ravets2013a}.

\begin{figure}[htb]
\centering
\includegraphics[width=0.6\linewidth]{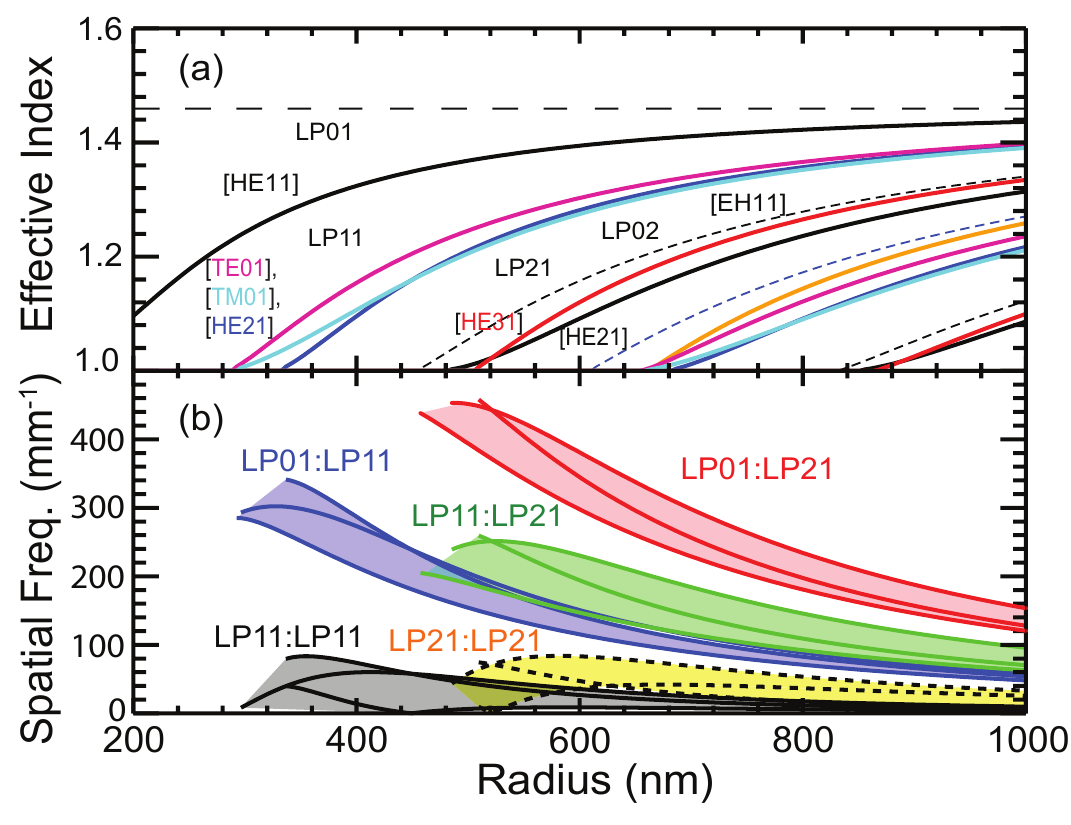}
\caption{Effective indices and spatial beating frequencies as a function of fiber radius for the lowest modes in an ONF. (a) Effective indices of several allowed modes in an ONF with $n_\mathrm{ONF}$ = 1.4534, surrounded by $n$ = 1.0, and a propagating wavelength of 795 nm.  The first four families are labeled with the specific modes in brackets and color-coded. 
(b)  Spatial frequencies between mode pairs for the first three mode families.  As a visual aid, all spatial frequencies between various mode families are colored the same way to show the groups of spatial frequencies observed at a particular $R$.  Of the 9 beat frequency curves between $LP_\mathrm{11}$ and $LP_\mathrm{21}$, only 3 are shown.
}
\label{fig:fig1}
\end{figure}

Each ONF mode has a waveguide propagation constant, $\beta$ = $kn_\mathrm{eff}$, where $k$ is the free-space propagation constant, 2$\pi/\lambda$. (See Fig.~\ref{fig:fig1} for the $R$ dependence of $n_{\rm{eff}}$ and the spatial frequencies between specific mode pairs). $n_\mathrm{eff}$ decreases with $R$, as the evanescent field samples more of the surrounding, lower index medium; when $n_\mathrm{eff}$ equals the surrounding index, the mode reaches cutoff and radiates out of the waveguide.  Fig.~\ref{fig:fig1}(a) shows $n_\mathrm{eff}$ for several modes in an ONF as a function of $R$.  Because of the high index contrast ($\Delta{n} \simeq 0.5$), several modes are allowed even when $R = \lambda$.  In this strong guiding regime, the scalar linearly-polarized (LP) basis commonly used for standard optical fibers is replaced by a full vector mode basis, but for convenience we still use the $LP$ basis to group the modes into families.  For example, the $TE_\mathrm{01}$, $TM_\mathrm{01}$, and two degenerate $HE_\mathrm{21}$ modes all belong to the $LP_\mathrm{11}$ mode family.  The $LP_\mathrm{01}$ family has two degenerate $HE_\mathrm{11}$ modes, so that between the $LP_\mathrm{01}$ and $LP_\mathrm{11}$ families, there are 6 unique beat frequencies.

Two modes with different $\beta_i$ and effective indices $n_\mathrm{i}$ will interfere with beat length 
$z_\mathrm{b} = 2\pi/({\beta_\mathrm{i}}-{\beta_\mathrm{j}})  = \lambda/(n_i-n_j)$, 
corresponding to a spatial frequency $\nu_\mathrm{b}=1/z_\mathrm{b} = (n_i - n_j)/\lambda$.  Figure 1(b) shows $\nu_b$ for several mode pairs in the lowest four mode families with groups of curves shaded according to which families are involved.  As $R$ decreases, the curves end abruptly when one of the modes reaches cutoff.  Near the cutoffs, interfamily beating is restricted to $z_b < 5~\mu$m ($\nu_b$> 200~mm$^{-1}$)~(e.g. between the $LP_{01}$ and $LP_{11}$ families), while for intrafamily beating, $z_b > 10$ $\mu$m ($\nu_b$< 100~mm$^{-1}$).  

\section{Experiment}\label{sec:Experiment}

The ONFs are drawn to design radii of 390 nm that support the $LP_{01}$ and $LP_{11}$ mode families.  We use a stabilized diode laser with $\lambda$ = 795 nm (Vescent D2-100-DBR) to launch superpositions of all 6 modes in the $LP_{01}$ and $LP_{11}$ families~\cite{Ravets2013a} so that beat frequencies between all possible pairs of propagating modes can be observed on the waist.

\begin{figure}[htb]
\centering
\includegraphics[width=0.6\linewidth]{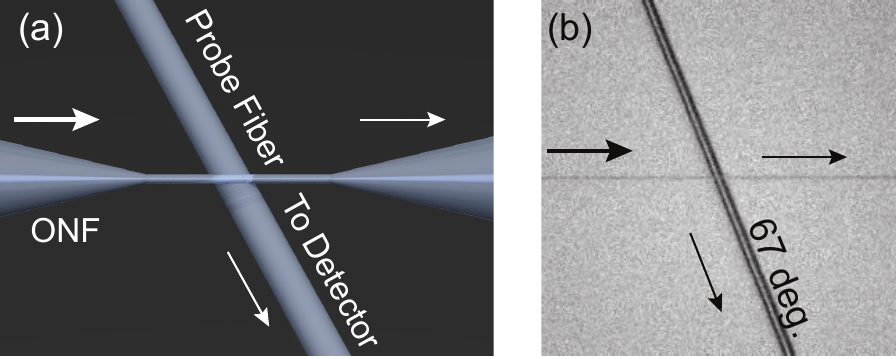}
\caption{(a)  Schematic of probe fiber with respect to ONF.  (b) Camera image of probe fiber and nanofiber on the ONF waist. }
\label{fig:fig2}
\end{figure}

To detect the propagating light, we use a fiber probe with 6-$\mu$m-diameter waist that contacts the ONF from below with a crossing angle of 67 degrees, as drawn in Fig.~\ref{fig:fig2}(a).  Figure~\ref{fig:fig2}(b) shows an image of the probe fiber in contact with the ONF.
The ONF has a taper angle $\Omega$ = 1 mrad, a design waist $R_w$ = 390 nm, and waist length $L$ = 10 mm.  At $R_w$ = 390 nm, the waist supports all modes in the $LP_{01}$ and $LP_{11}$ families at $\lambda$ = 795 nm.   A small crossing angle would provide excellent optical coupling to the probe fiber, but would decrease spatial resolution; a large crossing angle improves spatial resolution, but reduces optical coupling to propagating modes of the probe fiber.  Our crossing angle is a compromise between these issues and restrictions due to optical mounts and the apparatus.  The probe fiber is standard, single-mode SM750 optical fiber tapered in the same pulling apparatus.  Because this probe fiber is highly multimode in the waist, only a small percentage of the evanescently-coupled light reaches the detector, which measures only the light guided in the probe fiber core.  For our ONF, crossing angle, and probe fiber, the power detected is typically $10^{-5}-10^{-6}$ of the power propagating in the ONF waist (1--2 mW) for the modes of interest.  

The contact point is translated at 20 $\mu$m/sec over the ONF length by the same high-resolution stepper motors used during ONF fabrication to measure the propagation behavior in the ONF.  Although this approach could lead to scratches on the ONF~\cite{Sumetsky}, we monitored the transmitted power and intensity distribution before and after the measurement and saw no degradation. We have found that speeds slower than this can result in discontinuous signals due to stick-slip between the two fibers, but the ideal speed depends also on crossing angle and tension.

\section{Results}\label{sec:Results}
\subsection{Propagation behavior}\label{sec:rawdata}

\begin{figure}[htb]
\centering
\includegraphics[width=0.6\linewidth]{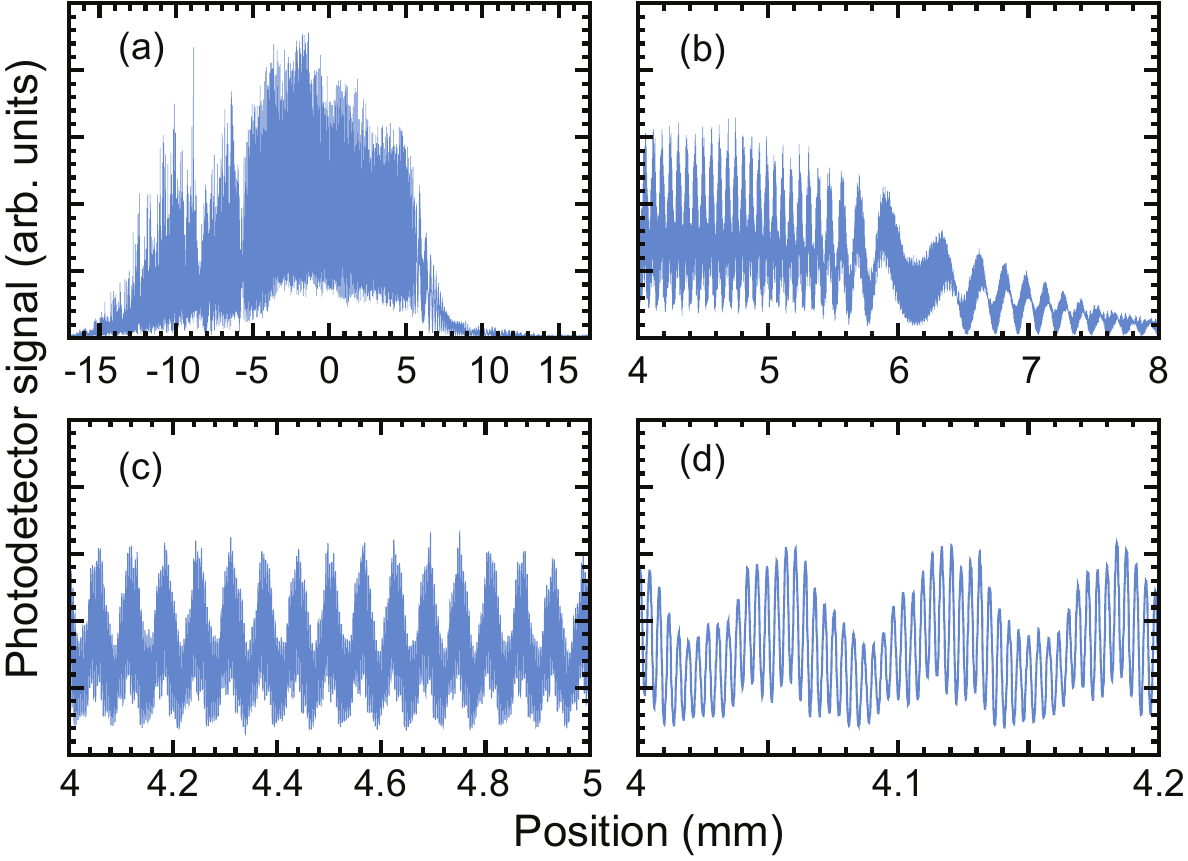}
\caption{(a) Near-field scanning probe signals over about 30 mm of the ONF. The waist is between $|z| \leq 5$ mm.  (b)-(d) Successive magnifications on the output side of the waist. 
}
\label{fig:fig3}
\end{figure}

Figure~\ref{fig:fig3} shows the detected light as a function of propagation distance ($z$) along the fiber.   Fig.~\ref{fig:fig3}(a) shows the detected power from about 15 mm before the center of the waist on the input taper, and continues to 15 mm onto the output taper.  The origin of the $x$-axis is the middle of the approximately 10-mm-long waist $L$.  We choose this origin because the taper is nominally symmetric, making comparisons between the modal distribution in the input and output tapers apparent.

Several properties of the light propagation can be observed in Fig.~\ref{fig:fig3}.  Because the field in the ONF is evanescently coupled to the probe fiber, the signal amplitude in Fig.~\ref{fig:fig3}(a) is largest on the waist ($|z| \leq 5$ mm).  The detected power is significantly greater and more heavily structured on the input taper ($z < -5$ mm) than on the output taper ($z > 5$ mm), where the average signal amplitude smoothly decays with clean oscillations.  This asymmetry is due to nonadiabaticity of the pull, causing excitation of higher-order modes when the light guidance transitions from core-cladding guidance to cladding-air guidance. On the input taper, these higher-order modes reach cutoff, eventually leaving only the $LP_{01}$ and $LP_{11}$ to interfere on the waist and output taper.  This leads to fewer observed beat frequencies, shown in the region from $z = 4-8$~mm of Fig.~\ref{fig:fig3}(b).  Figs.~\ref{fig:fig3}(b)-(d) show successively narrower ranges of data near the output of the waist, with two beat length scales apparent in Fig.~\ref{fig:fig3}(d).  The shorter length scale, with $z_b < 5 ~\mu$m, is due to interference between the $LP_{01}$ and $LP_{11}$ modes; the structure with $z_b \simeq 60~\mu$m is due to intramodal interference between members of the $LP_{11}$ family.

\subsection{Spectrogram analysis}\label{sec:spectrogram}

\begin{figure}[htb]
\centering
\includegraphics[width=0.6\linewidth]{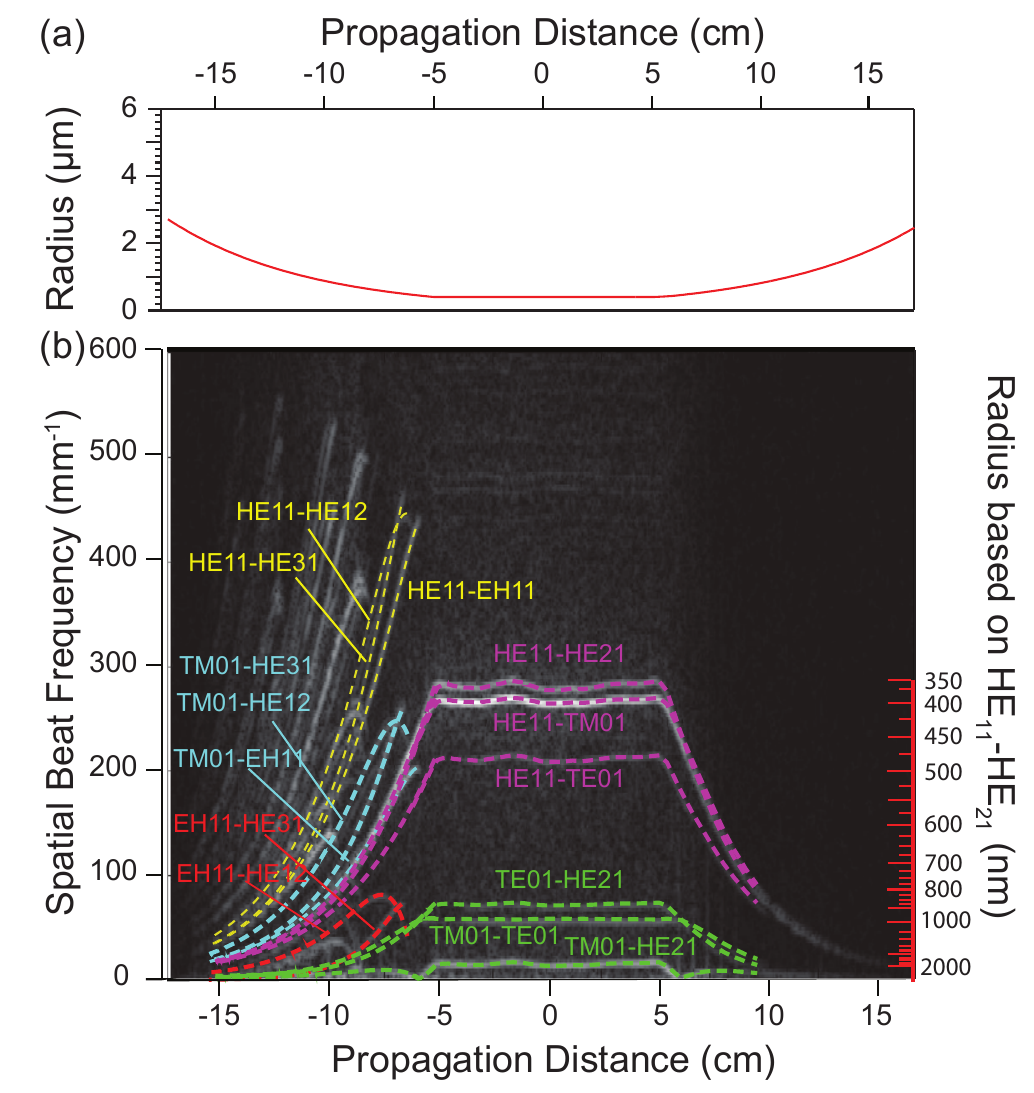}
\caption{(a) Design profile for ONF as a function of propagation distance.   (b) Spectrograms of the data, along with calculated beat-frequency curves for a number of mode pairs as a function of propagation distance.  Curves of the same color belong to the same family.  The right axis on (b) shows the extracted radius corresponding to the $HE_{11}:TM_{01}$ curve.}
\label{fig:fig4}
\end{figure}

We quantify the spatial frequencies and identify participating modes using spectrogram analysis~\cite{Ravets2013,Fatemi_optica}.  Briefly, we Fourier transform the signals measured in Fig.~\ref{fig:fig3}(a) with a sliding window of 600 $\mu$m width to calculate the local beat frequencies as a function of $z$.  Fig.~\ref{fig:fig4}(a) depicts the design profile for the fiber and Fig.~\ref{fig:fig4}(b) shows the resulting spectrogram.

Each curve in Fig.~\ref{fig:fig4}(b) is due to the interference of two modes.  On the input side ($z < -5$~mm), the complicated spatial structure observed in Fig.~\ref{fig:fig3}(a) is decomposed into numerous curves.  Far from the waist, each curve begins with both low amplitude and low spatial frequency.  The curves are less visible because in these regions most of the field is in the ONF, so the evanescent field is small.  Also, the effective indices far from cutoff are all approximately $n_{\rm{ONF}}$, so the beat frequencies are low.

Closer to the waist, the curve amplitudes and spatial frequencies increase.  Most of the curves abruptly end, similar to the calculations shown in Fig.~\ref{fig:fig1}(b).  
We show in Fig.~\ref{fig:fig4}(b) a number of calculated curves superimposed on the spectrogram to estimate $R$.  To label the unknown curves, we first determined the local $R$ as a function of $z$ using a single known spectrogram curve ($HE_{11}:TM_{01}$) and the calculations of Fig.~\ref{fig:fig1}(b).  The right axis shows this conversion. Using this $R(z)$, other mode pairs can be identified by comparison to Fig.~\ref{fig:fig4}(b).  The shortest visible beat lengths $z_{b}\simeq$~2.0 $\mu$m ($\nu_{b}\simeq 500$~mm$^{-1}$) are close to the shortest observable in fused silica corresponding to $z_{b}=2\lambda \simeq 1.6~\mu$m ($\nu_{b}\simeq$~625~mm$^{-1}$).

On the waist, only the fundamental $LP_{01}$ (one mode) and first higher order $LP_{11}$ (three modes) mode families can propagate, and the observed spatial frequencies are approximately constant.  With these two mode families, there are 6 observable and nondegenerate pairs of interfering modes on the waist (other faint features near 500 mm$^{-1}$ on the waist are due to aliasing artifacts).  The fundamental $HE_{11}$ mode interferes with the three nondegenerate modes of the $LP_{11}$ family with spatial frequencies near $\nu_b$=300~mm$^{-1}$.  Three additional frequencies below $\nu$~=~100~mm$^{-1}$ occur due to interference between the modes within the $LP_{11}$ family.  The lowest frequency curve, corresponding to the $TM_{01}:HE_{21}$ interference, shows a zero crossing at $|z| = 6$~mm, evident in the calculated curves shown in Fig.~\ref{fig:fig1}(b) near $R$ = 450~nm.

The observed $TE_{01}:TM_{01}$ interference is surprising, as these two modes are orthogonal everywhere and their individual modal powers are constant.  In the Rayleigh scattering results of Ref.\cite{Fatemi_optica}, we observed no beating between these two modes for this reason.  However, because our probe fiber is highly multimode at the point of contact and is single mode at the detector, the relative phase between the $TE_{01}$ and $TM_{01}$ modes that are evanescently coupled may affect the amount of light that reaches the detector, where the fiber is single mode, even though the strength of the evanescent field does not change.

\subsection{Fiber radius}\label{sec:radius}

\begin{figure}[htb]
\centering
\includegraphics[width=0.6\linewidth]{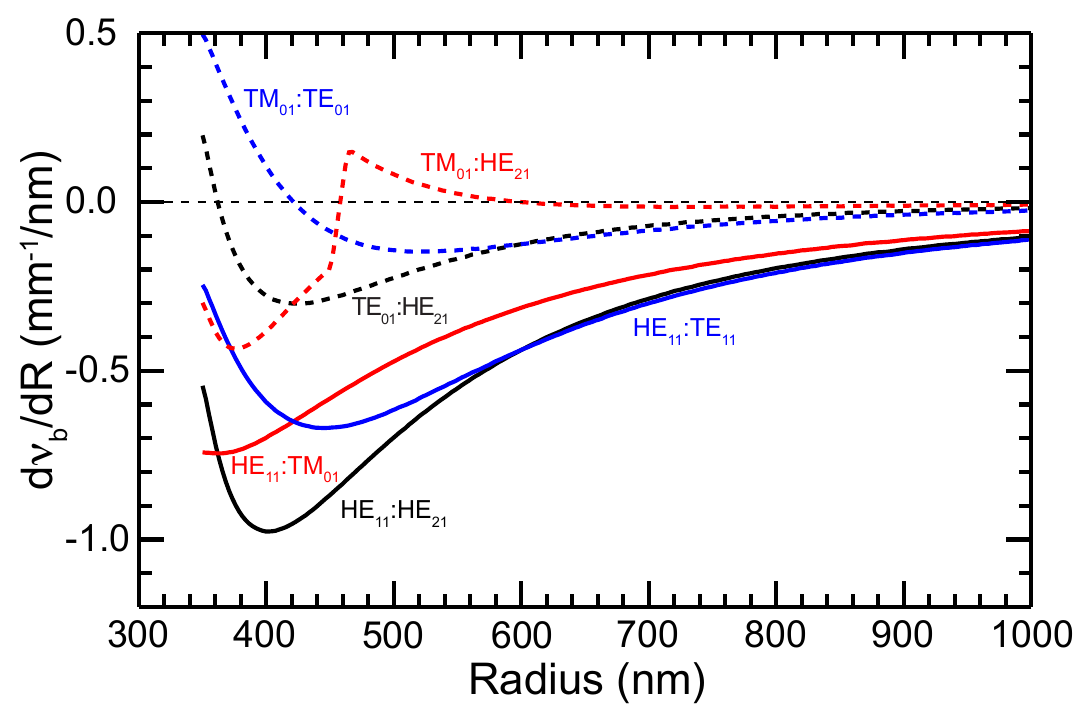}
\caption{Rate of change of the beat frequency with fiber radius. Note that the $HE_{11}:HE_{21}$, $HE_{11}:TE_{11}$, and $HE_{11}:TM_{01}$ have high sensitivity to radius near 400 nm.}
\label{fig:fig5}
\end{figure}

The beat length varies particularly steeply with $R$ near the mode cutoffs [Fig.~\ref{fig:fig1}(b) and Fig.~\ref{fig:fig4}(b)].  This property can be used to determine $R(z)$ with high precision. Fig.~\ref{fig:fig5} shows $d\nu_b/dR$ for each of the allowed mode pairs on the waist, at $\lambda=$795~nm.  The sensitivity is particularly strong for the $HE_{11}:LP_{01}$ curves (solid lines), with the $HE_{11}:HE_{21}$ curve achieving $d\nu_b/dR$= -1 mm$^{-1}$/nm, and the $HE_{11}:TM_{01}$ curve achieving $d\nu_b/dR$= -0.7 mm$^{-1}$/nm at $R$ = 400~nm.  Conversely, $z_b$ is insensitive to $R$ for the $TM_{01}:TE_{01}$ pair near $R$=420 nm, and the $TE_{01}:HE_{21}$ pair near $R=$360~nm.  The spectrogram in Fig.~\ref{fig:fig4}(b) shows these characteristics clearly in the waist region of this ONF, which had a design radius of $R$ = 390~nm:  The observed fluctuations in beat frequency are strongest for the $HE_{11}:HE_{21}$ curve, while the $TM_{01}:TE_{01}$ curve is constant.  At larger $R$, other mode pairs can be used for high sensitivity.

\begin{figure}[htb]
\centering
\includegraphics[width=0.6\linewidth]{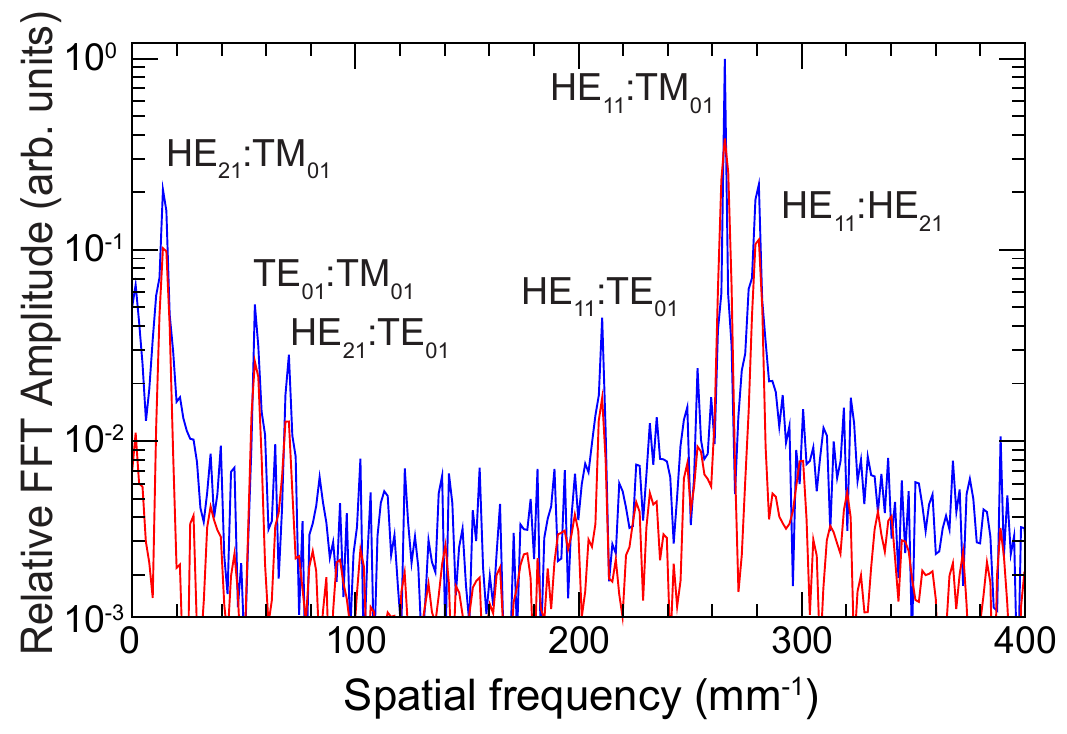}
\caption{FFT of a 600 $\mu$m long segment of the data  (waist of the ONF in Fig.~\ref{fig:fig4}). Note the logarithmic vertical scale. The blue trace is the raw data and the red has been multiplied by a Gaussian filter (see text for details).
}
\label{fig:fig5b}
\end{figure}

\begin{figure}[htb]
\centering
\includegraphics[width=0.6\linewidth]{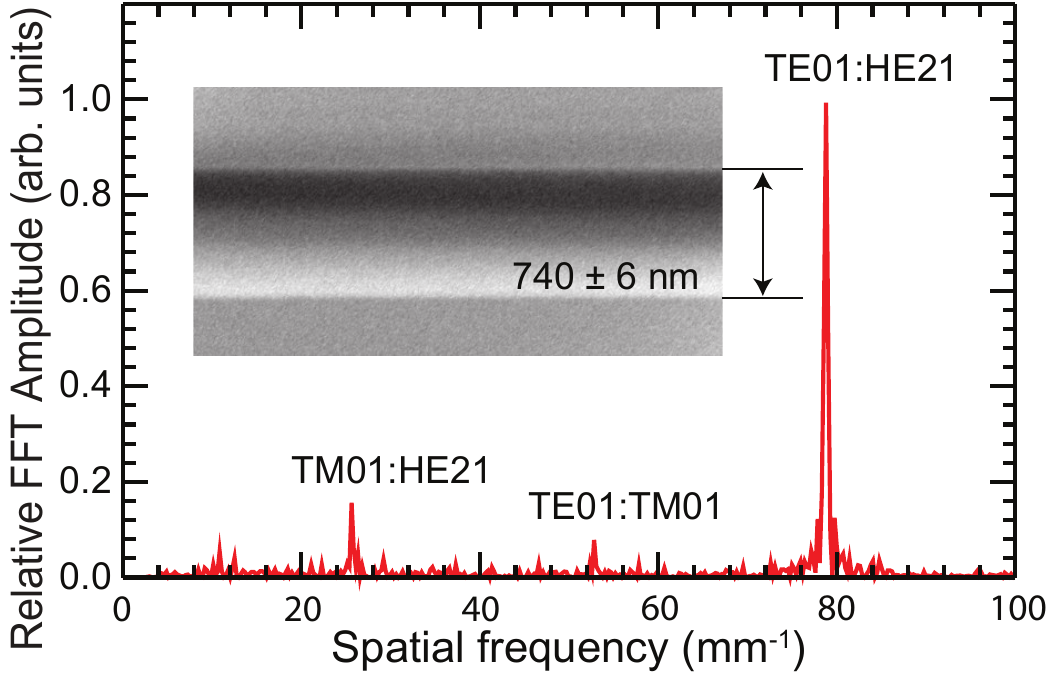}
\caption{ Beat frequencies for the fiber shown in the inset SEM image.  The low spatial frequency beat frequencies correspond to a fiber diameter of 742 $\pm$ 0.3 nm, in agreement with the SEM measurement of 740$\pm$6 nm.} 
\label{fig:fig6}
\end{figure}

We can find not only the value of $R$ at a given point but map its local variations with high precision on the waist.  To do this we determine the central frequencies of the features in each 600-$\mu$m-long FFT window comprising the spectrogram as in Fig.~\ref{fig:fig5b}.  Over this window size, which is smaller than the flame diameter used for heating the fiber, 
the beat frequencies are relatively constant and the spectral features are transform-limited.  Following the spectroscopic rule that the center of a resonance can be determined by its full width at half maximum  (FHHM) divided by the signal-to-noise ratio in the unfiltered data (blue trace of Fig.~\ref{fig:fig5b}), we estimate that the $HE_{11}:TM_{01}$ beat note near $265$ mm$^{-1}$ has a FWHM of 0.5 mm$^{-1}$ with a signal to noise better than 80. This estimate gives an uncertainty in the resonance center less than $0.01$~mm$^{-1}$, and an uncertainty in $R$ of $\approx0.01$~nm. 
The origin of the noise in Figs.~\ref{fig:fig5b}-\ref{fig:fig5c} is largely from stick-slip between the probe fiber and ONF, which causes phase discontinuities, and electronic noise.  Optical noise from bulk Rayleigh scattering inside the ONF has negligible coupling into the single mode of the probe fiber.

Having determined an estimate for the resolution of the method, we now proceed in a more quantitative way: The centers of transform-limited features in discrete FFTs can be accurately found by applying a mild Gaussian windowing function to the data  ($1/e^2$=400$~\mu$m) prior to calculating the FFT~\cite{Gasior} (red trace in Fig.~\ref{fig:fig5b}).  This broadens the spectral features to a known Gaussian form that can easily be fitted. Applying this approach to the $HE_{11}:TM_{01}$ curve, we find that the beat frequencies on the waist of the ONF vary from 271.13 $\pm$ 0.03 mm$^{-1}$ to 265.56 $\pm$ 0.03 mm$^{-1}$, corresponding to radii $R = 394.85 \pm 0.04$ nm to $R = 404.29 \pm 0.04$ nm over the  10 mm waist, where the uncertainty reflects 95\% confidence windows on the Gaussian fitting function using a 600-$\mu$m-long FFT window. 

We verified the accuracy of the method using scanning electron microscopy (SEM).  Fig.~\ref{fig:fig6} shows the lower part of the spectrum of spatial frequencies for another ONF that could then be destroyed in the SEM. 
Note that only two of the three marked beat frequencies are independent, with $\nu(TM_{01}:HE_{21})+\nu(TM_{01}:TE_{01})= \nu(TE_{01}:HE_{21)}$. The $TE_{01}:HE_{21}$ beat frequency, with signal-to-noise ratio close to 10, gives an estimate of the fiber radius of 742 $\pm$ 0.3 nm.  This value is within the bounds measured by the SEM of 740 $\pm$ 6 nm.  This fiber probe measurement is at least an order of magnitude more precise than SEM and is also nondestructive.

\subsection{Mode Control}\label{sec:control}
\begin{figure}[htb]
\centering
\includegraphics[width=0.6\linewidth]{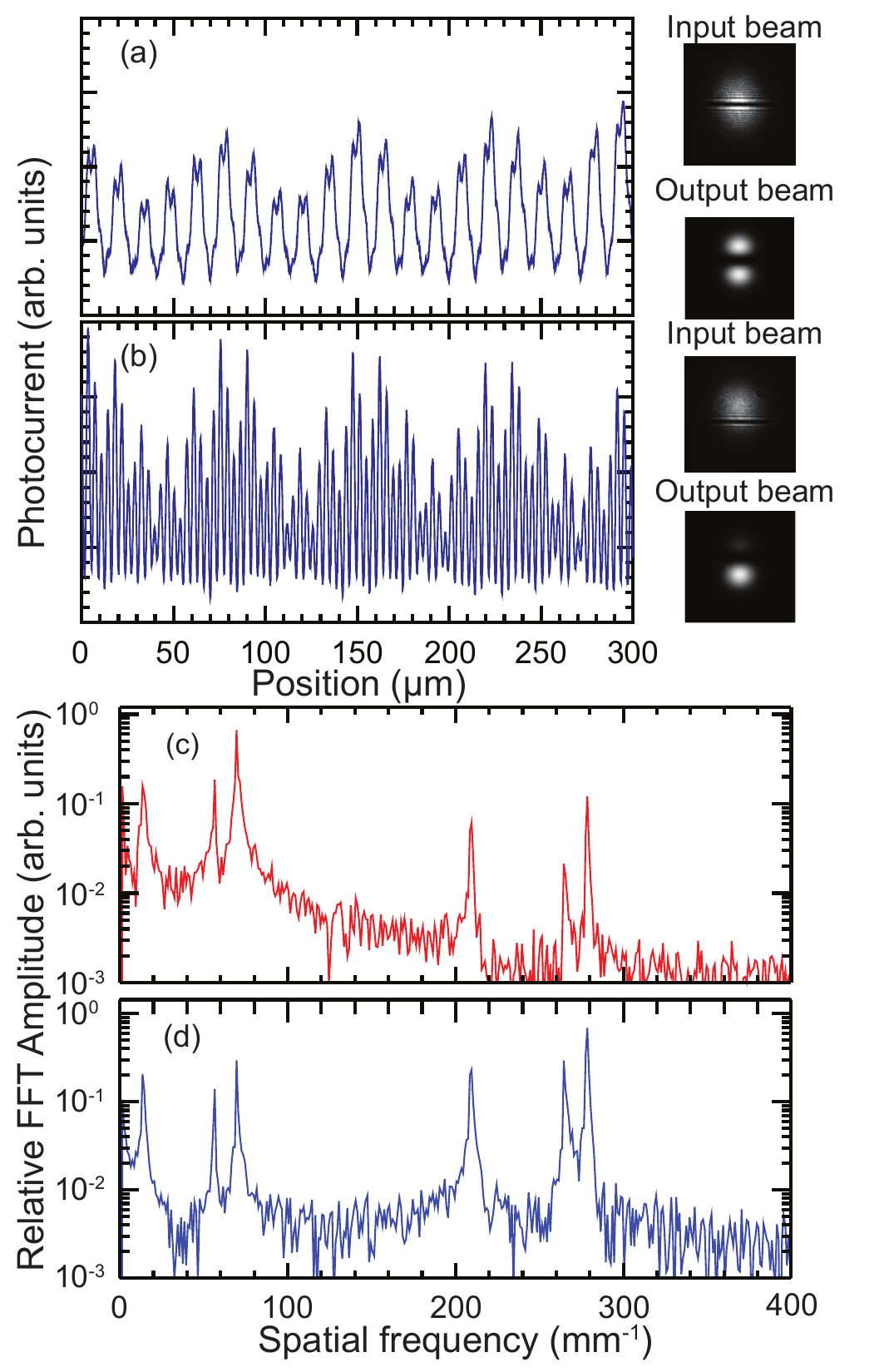}
\caption{Control of the modes beating at the fiber waist. (a) shows the spatial data when the $HE_{1,1}$ mode is suppressed. (b) shows the spatial data when the $HE_{11}$ is present.The two  photos to the right of (a) and (b) show the intensity profiles at the fiber input (top) and output (bottom) for the appropriate modes. (c) and (d) show the FFT of the 300 $\mu$m long spatial scan at the  waist of the ONF. Note the logarithmic vertical scale.   The dark line in the input beam pictures is caused by the $\pi$ phase jump on the phase plate.}
\label{fig:fig5c}
\end{figure}
This evanescent probe technique can spatially resolve high frequency interfamily mode beating. The amplitude of the beating is proportional to the product of two mode amplitudes. We can control the relative strengths of any pair of desired modes on the waist by looking at the magnitude of their beat note in the Fourier transform.  We demonstrate one use of the technique by suppressing the fundamental mode family propagating on the ONF waist in Fig.~\ref{fig:fig5c}. 
Relative weights of the $LP_{01}$ and $LP_{11}$ families are controlled by inserting a $\pi$-phase-plate into the input Gaussian beam~\cite{Ravets2013}.  Without a phase plate, well-aligned fiber coupling of the input Gaussian beam has maximum overlap with the $LP_{01}$ mode, whereas a centered plate ideally has only $LP_{11}$ contribution.  Mixtures of these families are produced by decentering the plate [intensity distributions at the fiber input and output are next to Fig.~\ref{fig:fig5c}(a) and Fig.~\ref{fig:fig5c}(b)]. Figures~\ref{fig:fig5c}(a) and Fig.~\ref{fig:fig5c}(b) show two traces over the same section of the ONF, with their respective FFTs on the bottom [Fig.~\ref{fig:fig5c}(c) and Fig.~\ref{fig:fig5c}(d)], when a phase plate is used to control the relative amount of $LP_{11}$ and $LP_{01}$ modal content within the ONF.  Fig.~\ref{fig:fig5c}(a) is the raw oscilloscope data when the fundamental mode is suppressed by centering a pi-phase-plate in the input beam. The peaks above 200~mm$^{-1}$, due to the $LP_{01}:LP_{11}$ beating, are smaller than than the intramodal $LP_{11}:LP_{11}$ beating.  When the phase plate is shifted down [input picture next to Fig.~\ref{fig:fig5c}(b)], so that the beam has a mixture of fundamental and higher modes, all 6 peaks in the FFT are strong. The highest beat frequencies at almost 300 mm$^{-1}$ correspond to beating with the fundamental $HE_{11}$ mode, producing oscillations with $z_{b} \approx 3.57~\mu$m, close to the theoretical limit of 2$\lambda$ for silica ONFs, something impossible to resolve with Rayleigh scattering~\cite{Fatemi_optica}.
 
\section{Discussion and Conclusions}

As $R$ increases, numerous inter- and intrafamily spatial frequency can be observed and identified using this near field scanning approach with sub-micron beat-length resolution.  This resolution is high enough to measure the minimum possible $z_b = \lambda/(n_{\rm{ONF}}-1)$ $\simeq 2\lambda = 1.6~\mu$m for this work with $n$ = 1.4534 and our propagating $\lambda$ = 795~nm.  The main limitation of the method comes from the requirement of an allowed higher-order mode family to provide measurable beat frequencies.  Single-mode operation occurs when $R \leq 0.363\lambda$ (V-number = 2.405), making measurements of $R < 100$~nm impractical because of the short optical wavelengths required. 

Good agreement with all beat frequency curves can only be satisfied for the correct value of $n$. 
We used the experimentally-obtained frequencies for one of the beat frequencies $(HE_{11}:TE_{01})$ to extract $R(z)$ based on that curve.  We then used these radii to calculate the 5 other curves.  An incorrect value of $n$ results in poor agreement; one cannot compensate for an incorrect value of $n$ by adjusting $R$.  For a fixed value of $R$ and a set of beat frequencies with uncertainties, as in Fig.~\ref{fig:fig5b}, of 0.03 mm$^{-1}$, the data are consistent only as long as the index agrees to within 2 x 10$^{-4}$~\cite{Malitson}.  The dispersion curve of fused silica is typically quoted to an accuracy 1 x 10$^{-5}$. This self-consistency check strongly suppresses any systematic effects in the procedures to either extract the radius or study the propagation of the higher order modes in the ONF.

The technique presented here has a high signal to noise ratio, which allows high spatial resolution. It does this with immunity to scattered light and ONF mechanical stress and vibrations. The mode-dependent coupling from the ONF to the probe fiber conditions the amplitudes of the beat resonances, but does not change the frequency of the beating. The amplitudes then give only qualitative information about the coupling. The average radius with its sub-nanometer resolution comes from the beating frequency, and shows robustness and repeatability while the scanning {\it in situ} is fast. The analysis of the transmission properties for higher order modes and their cutoffs give unique protection against systematic errors. The excellent spatial resolution in the radius determination can aid in evaluating the uniformity of the pulled fiber (\textit{e.g.} the noticeable variation in $R$ in the test fiber of Fig.~\ref{fig:fig3}, and can be used to optimize ONF production. 

\section*{Acknowledgements}
We thank Je-Hyung Kim for his help with the SEM imaging.  
\section*{Funding Information}
Army Research Office (ARO) (Atomtronics MURI (528418)); Defense Advanced Research Projects Agency (DARPA) (HR0011411122); National Science Foundation (NSF) (PHY-1430094); Office of Naval Research (ONR).

\bibliography{TaperFiberProbe}

\end{document}